\documentclass[%
 reprint,
 amsmath,amssymb,
 aps,
]{revtex4-2}

\usepackage{graphicx}
\usepackage{dcolumn}
\usepackage{bm}
\usepackage{circuitikz} 
\usepackage{subcaption}
\usepackage{svg}
\usepackage{placeins}
\usepackage{lipsum} 
\usepackage[colorlinks=true, urlcolor=blue, citecolor=blue]{hyperref}
\usepackage{url}  

\begin{document}

\preprint{APS/123-QED}

\title{Protected Symmetrical Superconducting Qubit Based on Quantum Flux Parametron}

\author{Kian Rafati Sajedi}
\email{kianrafaty@gmail.com}
\author{Mojtaba Hosseinpour Choubi}%
 \email{mojtabah72@yahoo.com}
\author{Mehdi Fardmanesh}
  \email{fardmanesh@sharif.edu}
  \homepage{\\http://serl.sharif.ir}
\affiliation{%
Superconductor Electronics Research Laboratory, Department of Electrical Engineering, Sharif University of Technology, 11155, Tehran, Iran
}%




\date{\today}

\begin{abstract}
  Conventional Quantum Flux Parametrons (QFPs) have been historically used for storing classical bits in Josephson junction-based
  computers. In this work, we propose a novel QFP-based topology dubbed “Degenerium” qubit, to process and compute quantum information.
  Degenerium combines principles from the 0-$\pi$ qubit and flux qubit to create ideally degenerate quantum ground states, while significantly
  simplifying the 0-$\pi$ qubit structure. The symmetrical design of Degenerium enables easier qubit control and fabrication. We demonstrate that
  due to the inherent symmetry of Degenerium, our designed qubit is insensitive to fabrication-induced variations in critical current ($I_c$) of the Josephson junctions.
  Our calculations of depolarization and dephasing rates due to charge, flux, and critical current noise
  sources result in depolarization and dephasing times of 1.25s and $90\mu s$, respectively. Further parameter tuning and optimization is possible to meet
  specific application demands.
\end{abstract}

\maketitle


\section{\label{sec:lntro}Introduction}

Classical digital computers have dominated information processing hardware for over half a century.
Since the invention of the transistor and the emergence of Moore’s law, computational capabilities have grown exponentially.
However, in recent years, the industry has approached its physical limits, with transistors shrinking to nanoscale dimensions \cite{waldrop2016chips}.

It was believed that Josephson-based computers were the follow-up technology after transistors, for their; (1) extremely fast 
switching speed, (2) extremely low power dissipation, and (3) operation at very low temperatures (around 4K).
Josephson junctions employ switching speeds less than 10ps \cite{Goto1986}.
Low power dissipation is attractive because this feature allows 
transfer of heat directly from chips into the liquid coolant bath. Additionally, Josephson junction circuits have the power dissipation 
in the order of 500nW per cycle \cite{likharev1991rsfq}. 
However this technology is dominated by the semiconductor industry due to high cost of operation of superconducting circuits.

Josephson junctions
also appeared as non-linear inductor elements in superconducting resonating circuits to construct quantum bits (qubits), i.e. "artificial atoms". This non-linear behavior allowed
quantum engineers to raise the anharmonicity of their qubits to isolate $|0\rangle$ and $|1\rangle$ states from higher-energy states, thus enabling the control and readout process \cite{Krantz2019}.
Now the focus of quantum engineers is to scale qubits and run quantum algorithms which take too long for any classical supercomputer \cite{NielsenChuang2010,Wang2012,StancilByrd2022}.

Building scalable and high-fidelity quantum computers has proved a challenging matter due to decoherence. Decoherence disturbs the state of the qubits
which stems from the coupling of qubits to their environment causing them to experience dephasing and depolarization that manifest themselves as errors in the computation,
making quantum algorithms likely to fail. It is hoped that eventually by constructing fault-tolerant quantum chips we will overcome this hurdle.

To this end,
the concept of symmetrically protected qubits have been developed \cite{Kitaev2003, Kitaev2006, Dennis2002, Gladchenko2009}. This concept allows one to define their states on near-degenerate ground states of the potential
rather than following the conventional energy-level-based qubits \cite{Manucharyan2009,Koch2007}.
Recently, it has been shown that symmetry in circuit topology 
can mitigate decoherence \cite{Brookes2022}.

In this work, we propose a novel qubit, which we call "Degenerium" \cite{RafatiSajedi2025}, that combines the classical use of the Quantum Flux Parametron (QFP) in Josephson computers with circuit topological symmetries,
drawing inspiration from the fault-tolerant design principles of Kitaev's 0–$\pi$ qubit. \cite{Kitaev2006,Brooks2013}

The structure of this paper is organized of sections and appendixes as follows; discussions on Degenerium's structure and Hamiltonian are in Sec.~\ref{sec:Degenerium}, where we talk about 
the schematic of Degenerium, its Hamiltonian derivation (further elaborated in Appendix.~\ref{app:Hamiltonian}), reasons for choice of parameters, and its classical potential.

In Sec.~\ref{sec:Dephasing}, we calculate the dephasing rates of Degenerium due to various sources of fluctuations.
Appendix.~\ref{app:Dephasing}, delves into how these rates have been calculated.
Appendix.~\ref{app:Depolarization}, elaborates how Hamiltonian is calculated considering charge noise.
In Sec.~\ref{sec:Depolarization}, depolarization rates are calculated for charge, flux and critical current noise sources.

\newcommand{\JJ}[5]{
    \draw (#1,#2) -- (#3,#4); 
    \pgfmathsetmacro{\midX}{(#1+#3)/2}
    \pgfmathsetmacro{\midY}{(#2+#4)/2}
    \draw (\midX-0.2,\midY+0.2) -- (\midX+0.2,\midY-0.2); 
    \draw (\midX-0.2,\midY-0.2) -- (\midX+0.2,\midY+0.2); 
    \node[anchor=south] at (\midX+0.5,\midY-0.3) {\(#5\)}; 
}

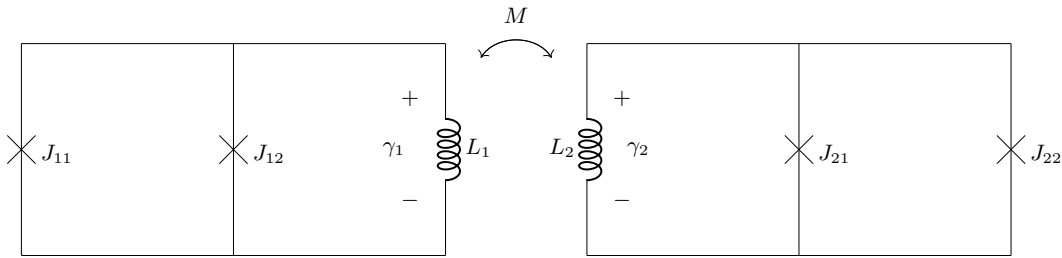
\begin{figure*}
  \centering
  \resizebox{0.8\textwidth}{!}{ 
  \begin{circuitikz}

    \JJ{0}{0}{0}{-3}{J_{11}}  
    \JJ{3}{0}{3}{-3}{J_{12}}  
    \draw (6,0) to[L, l=$L_1$] (6,-3);
    \draw (0,0) -- (6,0);
    \draw (0,-3) -- (6,-3);

    \draw (5.5,-1) node[above] {$+$};
    \draw (5.5,-2) node[below] {$-$};
    \draw (5,-1.5) node[right] {$\gamma_1$};

    \def\xstart{8} 

    \draw (\xstart,0) to[L, l_=$L_2$] (\xstart,-3); 
    \JJ{\xstart+3}{0}{\xstart+3}{-3}{J_{21}}  
    \JJ{\xstart+6}{0}{\xstart+6}{-3}{J_{22}}  
    \draw (\xstart,0) -- (\xstart+6,0);
    \draw (\xstart,-3) -- (\xstart+6,-3);

    \draw (\xstart+0.5,-1) node[above] {$+$};
    \draw (\xstart+0.5,-2) node[below] {$-$};
    \draw (\xstart+1,-1.5) node[left] {$\gamma_2$};

    \draw[->, thin] (6.5,-0.2) to[out=60, in=120] (7.5,-0.2);
    \draw[->, thin] (7.5,-0.2) to[out=120, in=60] (6.5,-0.2);
    \node at (7,0.4) {$M$};

\end{circuitikz}

  }
  \caption{Schematic for Degenerium qubit, which is composed of two QFPs, and $J_{i1}$ and $J_{i2}$ are ideally identical. 
  $M$ is the mutual inductance between the two inductors of QFPs, and $\gamma_1$ and $\gamma_2$ are gauge-invariant
  phases across $L_1$ and $L_2$, respectively.}
  \label{fig:circuit}
\end{figure*}

\section{\label{sec:Degenerium}Degenerium Qubit}

To store quantum information and process them, Degenerium leverages the circuit of QFP. It has
been shown in \cite{Takeuchi2013} that QFPs can operate in adiabatic processes which make them a 
great candidate for creating degenerate states for a quantum bit. To this end, we utilize the QFP circuit topology
to a more symmetrical structure that enables us to support degenerate quantum states within it.

The Degenerium circuit is shown in Fig 1, which is composed of two QFPs that are mutually coupled to each other via inductance. 
These QFPs are constructed by shunting an inductor to a Direct-Current Superconducting Quantum Interferance Device (DC SQUID).
By inducing fluxes within these SQUIDs we can modulate junction energy, $E_J$, of the junction pair on each side. This feature can be desirable
if one needs to adjust this energy by a magnetic flux rather than modifying the junctions through fabrication which will be inherently prone
to fabrication errors. 

In the classical QFP, an input current would create a persistent supercurrent
within either of its loops, each corresponding to a "0" or a "1" classical bit, respectively. This is the key principle of 
Rapid Single Flux Quantum (RSFQ) devices. In our setting, the logical qubit states are represented by distinct probability distributions of the gauge-invariant phase across the inductors in the circuit.

Additionally, Degenerium is biased by a flux line to create degenerate states for the qubit.
It will become clear, once we derive the potential of Degenerium, how biasing helps us achieve degeneracy.
The cornerstone of Degenerium are the loops $J_{12}$-$L_1$ and $J_{21}$-$L_2$ in Fig.1. If both of these loops are biased, Degenerium forms 
a quadruple-well potential, and if either of them are biased but not the other, it will form a double-well potential and if none are biased we will form a single-well potential
acting as a simple energy-level-based qubit. This condition of Degenerium allows for versatile connectivity on its quantum chip. If one loop is inaccessible by the flux line
we can bias the other loop.
This symmetrical structure helps us to avoid disturbances such as variations in junction fabrication.

To derive the Hamiltonian of the circuit we use the conventional method of circuit quantum electrodynamics (cQED)
much discussed in \cite{Devoret2017,Goto1986,Schuster2007}. We start by deriving the Lagrangian of the circuit

\begin{equation}
  \cal L = \cal T - \cal U
  \label{lagrangian}
\end{equation}

\noindent where $\cal T$ and $\cal U$ are the total kinetic energy and the total potential energy in the circuit, respectively. 
The exact calculations of this part are demonstrated in Appendix A, and here we show the key results of these calculations.
The kinetic energy and the potential are

\begin{align}
  \mathcal{T} &= \sum_{i=1}^{2} \frac{1}{2} C_{\Sigma i} \left( \dot{\Phi}_{J_{i1}}^2 + \dot{\Phi}_{J_{i2}}^2 \right) \label{eq:T}, \\
  \mathcal{U} &= \sum_{i=1}^{2} \frac{(\frac{\Phi_0}{2\pi})^2}{2L_i(1-K^2)} \gamma_i^2 \nonumber \\
  & \quad - \sum_{i=1}^{2} E_{Ji} \cos(\gamma_i - \phi_{bi} - \phi_i/2) \cos(\phi_i/2) \nonumber \\
  & \quad + \sum_{i=1}^{2} dE_{Ji} \sin(\gamma_i - \phi_{bi} - \phi_i/2) \sin(\phi_i/2) \nonumber \\
  & \quad + \frac{M (\frac{\Phi_0}{2\pi})^2}{L_1 L_2 (1 - K^2)} \gamma_1 \gamma_2, \label{eq:U}
\end{align}

\begin{figure*}
  \centering
  \begin{subfigure}[b]{0.32\textwidth}
    \includegraphics[width=\textwidth]{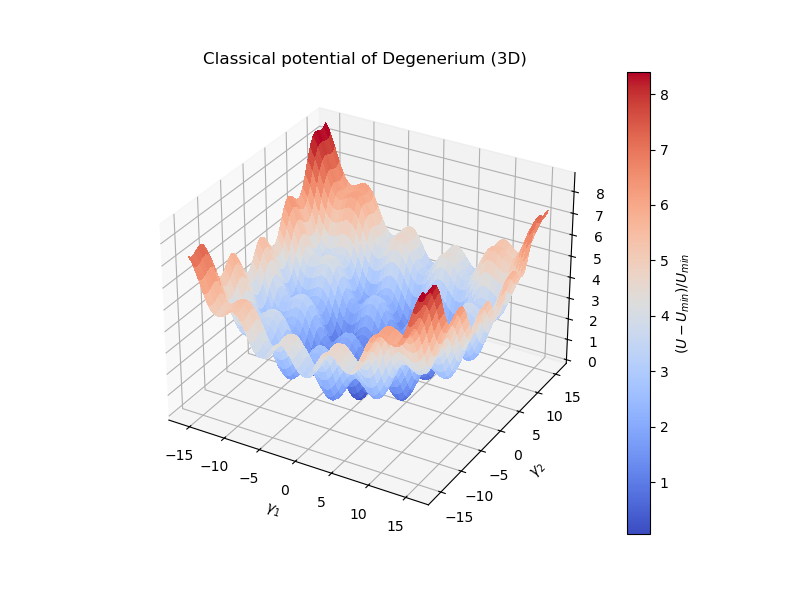}
    \caption{}
    \label{fig:sub1}
  \end{subfigure}
  \hfill
  \begin{subfigure}[b]{0.32\textwidth}
    \includegraphics[width=\textwidth]{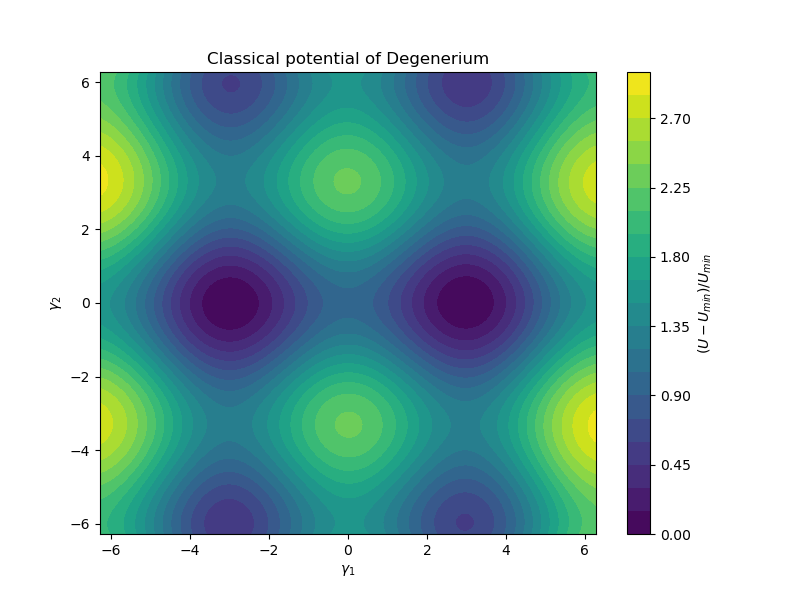}
    \caption{}
    \label{fig:sub2}
  \end{subfigure}
  \hfill
  \begin{subfigure}[b]{0.32\textwidth}
    \includegraphics[width=\textwidth]{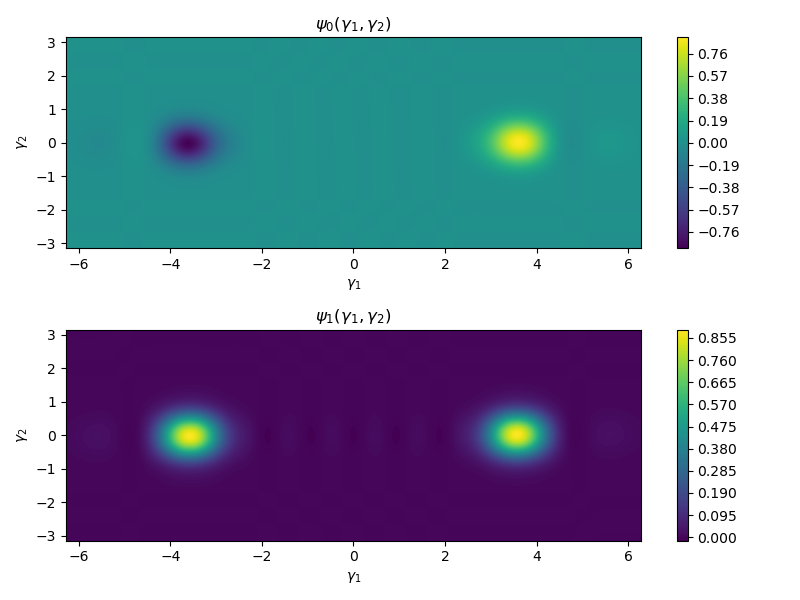}
    \caption{}
    \label{fig:sub3}
  \end{subfigure}
  \caption{: Degenerium’s a) 3D and b) 2D classical potential plotted against $\gamma_1$ and $\gamma_2$, the phases of the coupled inductors, 
  for $E_{Ci}/h=1 GHz$ , $E_{Ji}/h=100 GHz$ , $E_{Li}/h=10 GHz$ and $g/h=1GHz$.The bottom wells host near-degenerate states that we define our states upon.
  c) Wavefunctions for groundstate and its near-degenerate counterpart for Degenerium's $|0\rangle$ and $|1\rangle$ states.
  }
  \label{fig:energy_spectra}
\end{figure*}

\noindent where $E_{Ji}$ is the common-mode Josephson energy of each side (i.e. the common energy that both junctions on each side possesses) defined by $ \frac{\Phi_0}{2\pi} I_c$ ($\Phi_0 = \frac{h}{2e}$ is 
the superconducting flux quantum (or fluxon), which has a value of approximately $2.068 \times 10^{-15}\text{Wb}$), 
$dE_{Ji}$ is the differential junction energy of each side which captures variations in critical current of junctions due to fabrication errors,
$L_i$ and $M$ are the inductance of each side's inductor and the mutual inductance between them, respectively. 
We also define the coupling coefficient which modifies the effective inductance of each inductor as $K=\frac{M}{\sqrt{L_1L_2}}$. $C_{\Sigma i}$ is the total capacitance on
each side, $\Phi_{Ji1}$ and $\Phi_{Ji2}$ are fluxes threading in each side's first and second junctions, respectively.
Lastly, we define reduced fluxes $\gamma_i$, $\phi_{bi}$ and $\phi_i$ which are $2\pi \frac{\Phi_{Li}}{\Phi_0}$, $2\pi \frac{\Phi_{bi}}{\Phi_0}$ and $2\pi \frac{\Phi_i}{\Phi_0}$, respectively.
Where $\Phi_{Li}$ is the flux threading each side's inductor, $\Phi_{bi}$ is the biasing flux induced by an external flux line that threads either the $J_{21}$-$L_2$ or the $J_{12}$-$L_1$, and $\Phi_i$ is the flux threading the SQUID loop of each side modifying
their junction energy.

The classical potential of Degenerium is illustrated in Fig.2a and Fig.2b. It is evident that by operating in a $\gamma_1$ and $\gamma_2$ space,
we can form a potential that is composed of cosine and parabolic terms with their corresponding weights, i.e. $E_{Ji}$ and $E_{Li}=(\frac{\Phi_0}{2\pi})^2/(2L_i(1-K^2))$, respectively. By flux biasing, we can displace the potential wells
in the ($\gamma_1$,$\gamma_2$) space. For example the $J_{12}$-$L_1$ at $\Phi_{b1}=\Phi_0/2$ and leaving $J_{21}$-$L_2$ with no biasing flux
$\Phi_{b2}=0$, we can form a double-well potential in the $\gamma_1$ axis, Fig. 2.b. Similarly, biasing the $J_{21}$-$L_2$ at $\Phi_{b2}=\Phi_0/2$ and leaving $J_{12}$-$L_1$ with no biasing flux
$\Phi_{b1}=0$, can form a double-well potential in the $\gamma_2$ axis.

Additionally, by setting $E_{Ji}/E_{Li} \gg 1$, Degenerium enters a regime where its classical potential becomes
periodic in the $(\gamma_1, \gamma_2)$ space, resembling that of a conventional 0-$\pi$ qubit~\cite{Dempster2014}.
In this limit, the low-energy states can be approximated as superpositions of wavefunctions localized in alternating wells,
corresponding to logical $|0\rangle$ and $|1\rangle$ states distinguished by their parity symmetry.

To obtain the Hamiltonian from the Lagrangian, we first determine the conjugate momentum of our circuit and by
applying a Legendre transformation, we derive the classical form. Finally, by quantization,
we arrive at the complete Hamiltonian in the form

\begin{align}
  \hat{H} &= \sum_{i=1}^{2} 4 E_{C_{\Sigma i}} \hat{n}_i^2 + \sum_{i=1}^{2} \frac{1}{2} E_{L_i} \hat{\gamma}_i^2 \nonumber \\
  & \quad - \sum_{i=1}^{2} 2 E_{J_i} \cos\left( \hat{\gamma}_i - \phi_{b_i} - \frac{\phi_i}{2} \right) \cos\left( \frac{\phi_i}{2} \right) \nonumber \\
  & \quad + \sum_{i=1}^{2} 2 d E_{J_i} \sin\left( \hat{\gamma}_i - \phi_{b_i} - \frac{\phi_i}{2} \right) \sin\left( \frac{\phi_i}{2} \right) \nonumber \\
  & \quad - g \hat{\gamma}_1 \hat{\gamma}_2 \label{eq:H}
\end{align}

where $E_{C_{\Sigma_i}} = e^2/(2 C_{\Sigma_i})$ is the total capacitance energy of each side and $g = (\Phi_0/2\pi)^2 \cdot  M/(L_1 L_2 (1 - K^2))$ is the interaction term between the two sides.
$\hat{n}_i$ is the number of cooper pairs operator on each side.

Following our previous discussion on Degenerium's potential wells, we define our $|0\rangle$ and $|1\rangle$ states on the ground states within these potential wells.
We have illustrated these two states for $\Phi_{b1}=\Phi_0/2$ and $\Phi_{b2}=0$ in Fig.3c. As evident from this figure, by forming a double-potential well in the $\gamma_1$ axis
we have constructed two delocalized wavefunctions with respect to $\gamma_2$. Conversely, delocalized states can form with respect to $\gamma_1$ if we bias
oppositely, suggesting that higher fidelity can be achieved by controlling one side of the qubit and
reading out the state from the other side.

We can switch to 0-$\pi$ qubit operating regime by setting $E_{Ji}/E_{Li} \gg 1$ which forms a periodic potential well which is suitable for defining states in odd and even potential wells \cite{Dempster2014}. 
To achieve this configuration a very large inductance is needed as mentioned in \cite{Kitaev2006}.

\begin{figure*}
  \centering
  \begin{subfigure}[b]{0.32\textwidth}
      \includegraphics[width=\textwidth]{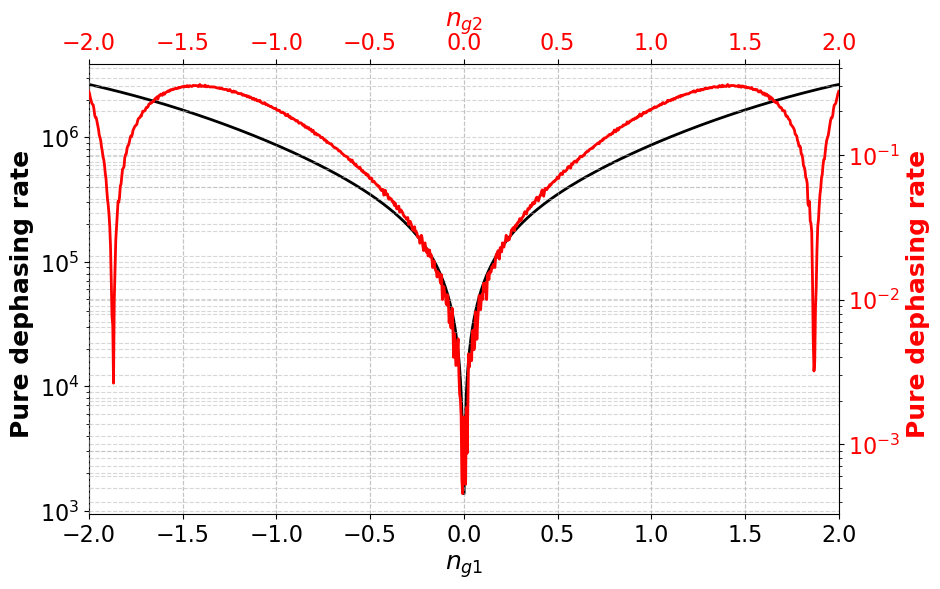}
      \caption{}
      \label{fig:sub1}
  \end{subfigure}
  \hfill
  \begin{subfigure}[b]{0.32\textwidth}
      \includegraphics[width=\textwidth]{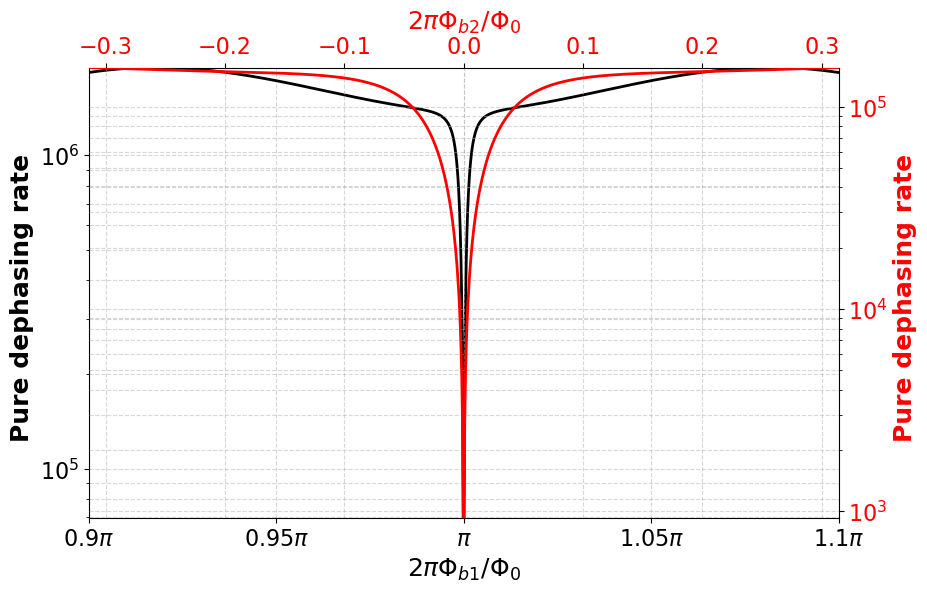}
      \caption{}
      \label{fig:sub2}
  \end{subfigure}
  \hfill
  \begin{subfigure}[b]{0.32\textwidth}
      \includegraphics[width=\textwidth]{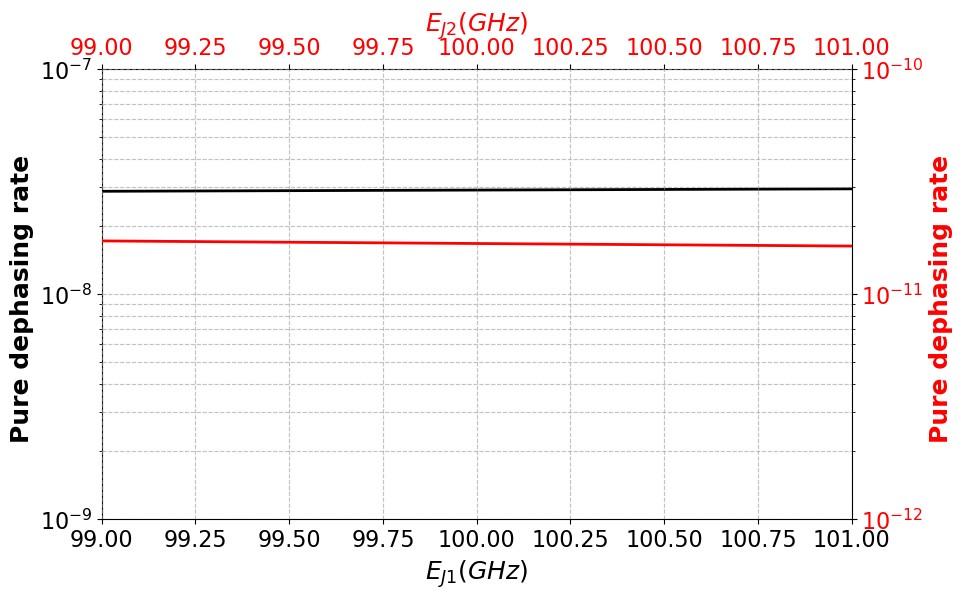}
      \caption{}
      \label{fig:sub3}
  \end{subfigure}
  \caption{Calculated pure dephasing rate due to 1/f noise in a) $n_{g1}$(black), charge noise of biased loop,  and $n_{g2}$ (red), charge noise of unbiased loop, ,b) $\Phi_{b1}$ (black), bias flux noise of biased loop, and $\Phi_{b2}$ (red), bias flux of unbiased loop, and c) $E_{J1}$ (black), Josephson energy of biased loop junctions, and $E_{J2}$ (red), Josephson energy of unbiased loop junctions.}
  \label{dephasing rates}
\end{figure*}

\section{\label{sec:Dephasing}Dephasing Rates}

To obtain Degenerium coherence characteristics, we first need to determine its pure dephasing and depolarization rates ($\Gamma_\phi$ and $\Gamma_1$ respectively) according to Bloch-Redfield theorem \cite{bloch1957generalized,redfield1957theory}. In this section, we focus on calculating the pure dephasing rates of Degenerium due to multiple sources of noise channels including charge, bias flux, and critical current of Josephson junction (i.e. Josephson energy). Additionally, we assume that the noise sources are uncorrelated, thus the total pure dephasing rate is the sum of all noise channels that are coupled to the main Hamiltonian of Degenerium. 

First, for arbitrary noise channel defined by $X(t)=X_0 + \delta X(t)$ which is a stationary Gaussian random process, the spectral power density of $X(t)$ can be calculated as
\begin{equation}
  S_X(\omega)=\int_{-\infty}^\infty  \langle X(0)X(t)\rangle e^{-i\omega t} dt.
\end{equation}

The dominant spectral noise density in the superconducting qubits is 1/f (flicker) noise which its corresponding spectral power density \cite{Martinis2005} is described by Eq.(\ref{1/f}) for a range of frequencies limited between infrared ($\omega_{ir}$) and ultraviolet ($\omega_{c}$) cutoff frequencies \cite{Ithier2005}.

\begin{equation}
  S_X^{1/f}(\omega)=\frac{2\pi (A_X)^2}{|\omega|}   ,\; \omega_{ir}<|\omega|<\omega_c
  \label{1/f}
\end{equation}

Here $A_X$ is the spectral power noise amplitude. In the weak noise approximation, the perturbed Hamiltonian is expanded in terms of $X(t)$ to capture the noise effect in the $V_X$ operator.

\begin{equation}
\label{pert}
  \begin{aligned}
     \hat{H}(X)&\approx \hat{H}(X_0)+(\frac{\partial \hat{H}}{\partial X}|_{X=X_0} )\delta X(t)\\
    &+\frac{1}{2}(\frac{\partial^2 \hat{H}}{\partial X^2}|_{X=X_0} )\delta X^2(t) 
      =\hat{H}(X_0)+\hat{V}_X
  \end{aligned}
\end{equation}

Finally, the pure dephasing rate can be calculated using Eq.(\ref{phase}) by considering the
decay rate of off-diagonal terms in the Bloch-Redfield picture of the qubit density matrix, which is caused by noise 
effect ($\hat{V}_X$) in the main Hamiltonian of Degenerium (Appendix B) \cite{Ithier2005}.

\begin{multline}
  \label{phase}
  \Gamma_X^\phi = 
  \Bigg[
  2 A_X^2 \left( \left. \frac{\partial \omega_{01}}{\partial X} \right|_{X=X_0} \right)^2 
  \left| \ln \left( \omega_{ir} \tau \right) \right| \\
  + 2 A_X^4 \left( \left. \frac{\partial^2 \omega_{01}}{\partial X^2} \right|_{X=X_0} \right)^2 
  \left( \ln^2 \left( \frac{\omega_c}{\omega_{ir}} \right) + 2 \ln^2 \left( \omega_{ir} \tau \right) \right)
  \Bigg]^{1/2}
\end{multline}

where $\omega_{01}$ is the transition frequency from $|0\rangle$ to $|1\rangle$ state in the qubit and $\tau$ is the duaration of calculation. Then, for different noise channels, the corresponding pure dephasing rates are calculated.

\subsection{Charge Noise}

The charge noise is modeled as a voltage source ($V_g$) that is capacitively coupled to Degenerium's circuit
through $C_g$ which is estimated to be $0.05C_\Sigma$. The noisy circuit is depicted in Fig.\ref{fig:charge noise circuit}.

The Hamiltonian of the circuit in Fig.\ref{fig:charge noise circuit} is obtained by using a similar method
as in Section II (for exact derivation see Appendix C). Then, we arrive at 

\begin{equation}
\label{charge noise}
\begin{aligned}
\hat{H} = & 4 \sum_{i=1}^2 \bigl[ E_{C_i'}  (\hat n_i-n_{gi})^2 - E_{gi} n_{gi}^2 \bigr] \\
& + \tfrac{1}{2} \sum_{i=1}^2 E_{Li} \hat{\gamma_i}^2 - g \hat{\gamma_1} \hat{\gamma_2} \\
& - 2 \sum_{i=1}^2 E_{Ji} \cos(\hat{\gamma_i} - \phi_{bi} - \phi_i/2) \cos(\phi_i/2)
\end{aligned}
\end{equation}

where $n_{gi}=\frac{C_{gi}V_{gi}}{2e}$, $E_{C_i'} = \frac{e^2}{2(C_{\Sigma i} + C_{gi})}$, and $E_{gi} = \frac{e^2}{2 C_{gi}}$.
The charge noise pure dephasing rate, according to Eq.(\ref{phase}), is calculated by computing $\omega_{01}$ versus
the various amount of $n_g$. Then, we proceed to calculate the pure dephasing rate. The final simulation result is demonstrated in Fig.\ref{dephasing rates}a.

For simulations, we use python QuTip library \cite{johansson2012qutip} to calculate Hamiltonian's eigenvalues and eigenstates. In addition,
we assume $\omega_{ir}=1 Hz$, $\omega_c=3 GHz$ and $\tau=10\mu s$ \cite{yan2016flux}. Additionally, the charge spectral noise power amplitude is taken
to be $A_{n_g}=10^{-4} 1/\sqrt{Hz}$ \cite{zorin1996background}. This noise charge models pure dephasing rate due to qubit bulk charge defects
or quasiparticles \cite{Lutchyn2006,Oliver2013,Murray2021} that tunnel through the junction which for Degenerium, it changes from approximately $10^3$ rad/s at $n_{g1}=0$ to $10^6$ rad/s
at $n_{g1}=2$. This estimation will not hold for high values of $n_g$ due to weak noise approximation assumption. Furthermore, the biased
side of Degenerium circuit is more sensitive to charge noise than the unbiased side.

\begin{figure*}
  \centering
  \begin{subfigure}[b]{0.32\textwidth}
      \includegraphics[width=\textwidth]{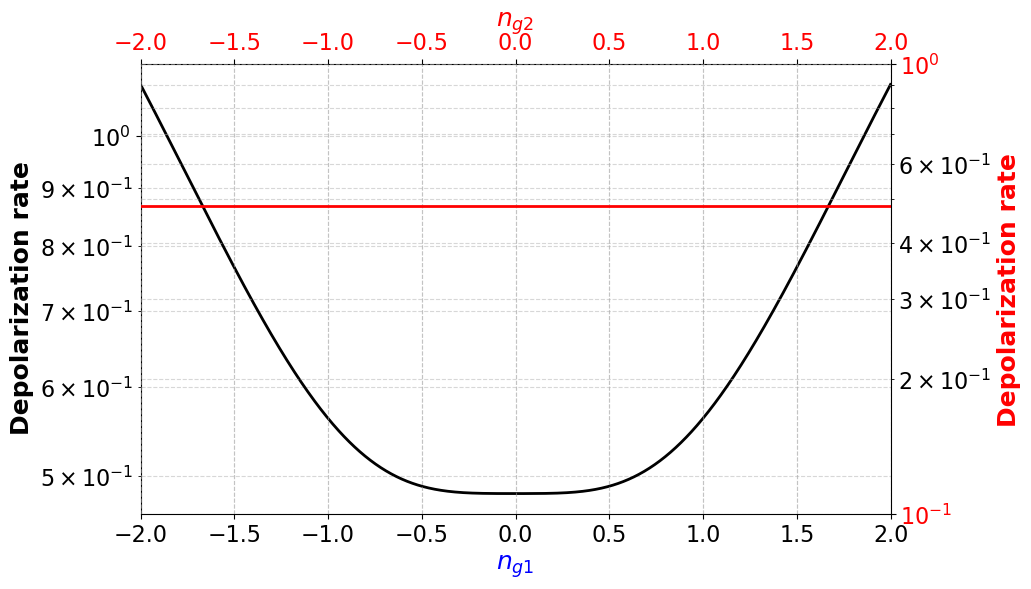}
      \caption{}
      \label{fig:sub1}
  \end{subfigure}
  \hfill
  \begin{subfigure}[b]{0.32\textwidth}
      \includegraphics[width=\textwidth]{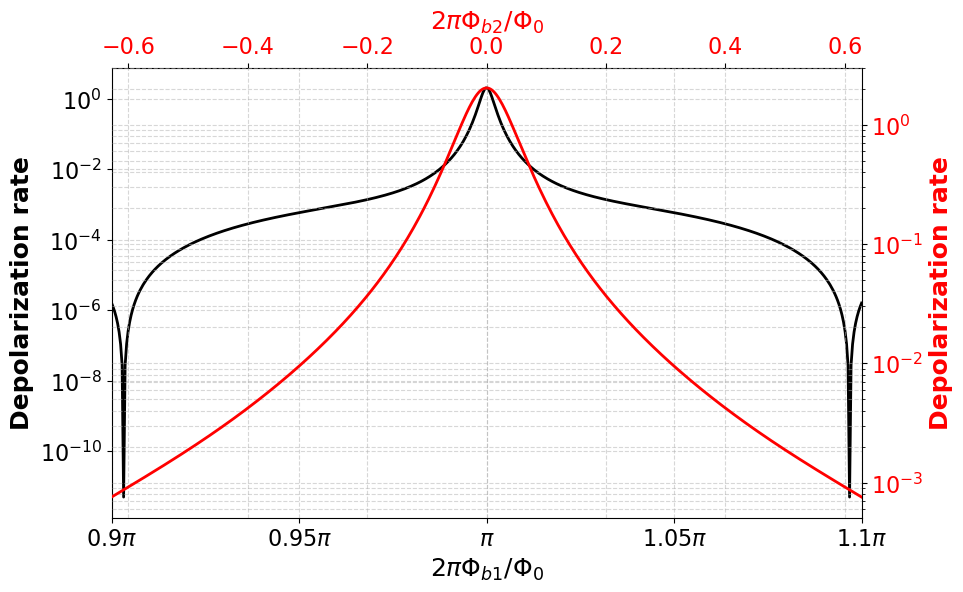}
      \caption{}
      \label{fig:sub2}
  \end{subfigure}
  \hfill
  \begin{subfigure}[b]{0.32\textwidth}
      \includegraphics[width=\textwidth]{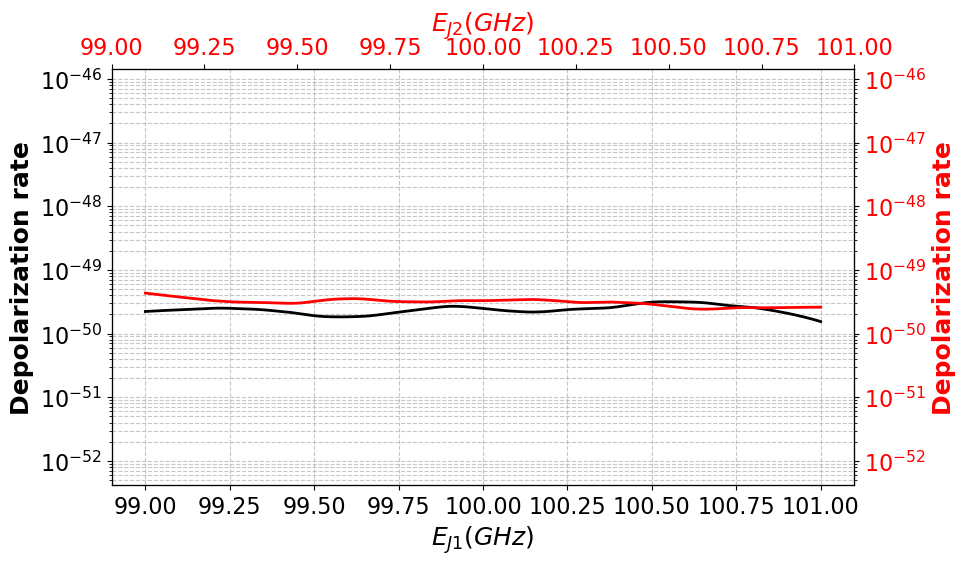}
      \caption{}
      \label{fig:sub3}
  \end{subfigure}
  \caption{Calculated depolarization rate due to 1/f noise in a) $n_{g1}$(black), charge noise of biased loop,  and $n_{g2}$ (red), charge noise of unbiased loop, ,b) $\Phi_{b1}$ (black), bias flux noise of biased loop, and $\Phi_{b2}$ (red), bias flux of unbiased loop, and c) $E_{J1}$ (black), Josephson energy of biased loop junctions, and $E_{J2}$ (red), Josephson energy of unbiased loop junctions.}
  \label{depolarization rates}
\end{figure*}
\begin{figure}[h]
    \centering
    \resizebox{0.5\textwidth}{!}{ 
    \begin{circuitikz}[scale=0.9] 
      \JJ{0}{0}{0}{-3}{J_{l1}}  
      \JJ{3}{0}{3}{-3}{J_{l2}}  
      \draw (6,0) to[L, l=$L_1$] (6,-3);
      \draw (0,0) -- (6,0);
      \draw (0,-3) -- (6,-3);
    
      \draw (5.5,-1) node[above] {$+$};
      \draw (5.5,-2) node[below] {$-$};
      \draw (5,-1.5) node[right] {$\gamma_1$};
    
      \draw (-2,0) to[american voltage source, l=$V_g$] (-2,-3); 
      \draw (-2,0) to[C, l=$C_g$] (0,0); 
      \draw (-2,-3) to (0,-3);
    
      \node at (0,0) [circle, fill, inner sep=1pt] {}; 
      \node at (0,-3) [circle, fill, inner sep=1pt] {}; 
      
      \draw(7.2,-1.5) node {$\cdots$};

      \draw[->, thin] (7,0.2) to[out=180, in=90] (6.3,-0.3);
      \node at (7,0.4) {$M$};
    \end{circuitikz}
    }
    \caption{Schematic for half of Degenerium's circuit that is capacitively coupled to a charge noise voltage.
    Similar configuration is setup on the other side.}
    \label{fig:charge noise circuit}
\end{figure}
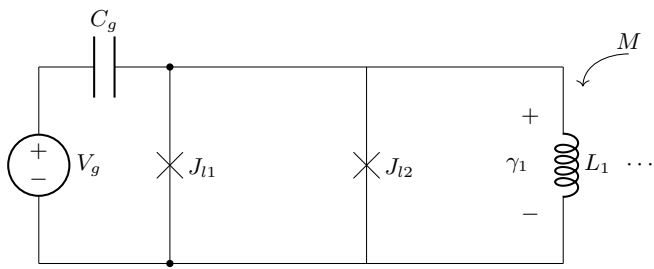

\subsection{Bias Flux Noise}

Similar to the previous section, the effect of 1/f noise in the bias flux is captured by simulation of the transition frequency of Degenerium versus fluctuations in the bias flux.

For the simulations, we assume the flux spectral noise power amplitude to be 
$A_{\Phi}=10^{-6} \Phi_0/\sqrt{Hz}$ which is a typical value for a 1/f flux noise
in superconducting qubits \cite{hutchings2017tunable}. All other parameters are
identical to the previous section. The resulting pure dephasing rate due 
to bias flux noise from simulations is depicted in Fig.\ref{dephasing rates}b.

The bias flux noise has the most impact on pure dephasing rate since the potential
is dependent solely on this channel. Moreover, any fluctuation in flux bias will degrade
the dephasing rate making the bias flux the main bottleneck of Degenerium. The flux noise
channel contributes $10^5$ rad/s rate to total pure dephasing rate. By improving bias flux
noise, the total operation of Degenerium will be enhanced significantly.

\subsection{Critical Current Noise}
The critical current of Josephson junction plays a pivotal role in the energy stored in the junction.
However, its effect on the transition frequency of Degenerium is negligible compared to other sources of
noise.

The pure dephasing rate due to noise in the Josephson junction energy ($E_J$) is demonstrated in
Fig.\ref{dephasing rates}c. The critical current spectral noise power amplitude is estimated to
be $A_{I_C}=10^{-7} I_C/\sqrt{Hz}$ \cite{Koch2007} and other parameters for simulation remained the same.

Degenerium is resilient against fluctuations in critical current of Josephson junctions (i.e. energies stored
in the junctions). Notice that these $E_{Ji}$ variations refer to common mode of each side's
junction pair. Additionally, pure dephasing rate due to critical current noise is estimated to be approximately
$10^{-10}$ rad/s which is negligible compared to other pure dephasing rates.

\section{\label{sec:Depolarization}Depolarization Rates}

Conventional qubits establish their states based on difference of energy level which will lead to small depolarization rates.
Depolarization occurs via different mechanisms such as spontaneous transitions between
energy eigenstates, coupling to the environment such as defects within the superconducting chip or other two-level-system (TLS) defects that the qubit can
couple to \cite{Martinis2005,Oliver2013}. In this section we will discuss the depolarization rate of Degenerium. 

We characterize this decoherence time scale by the parameter $T_1$ which is called the depolarization time. Its reciprocal $\Gamma_1$, the depolarization rate,
is given by the sum $\Gamma_1 = \Gamma_{0\to1} + \Gamma_{1\to0}$, where $\Gamma_{1\to0}$ is the relaxation rate and $\Gamma_{0\to1}$ is the excitation rate. First we define
the operator coupling Degenerium's degrees of freedom  $\gamma_1$ and $\gamma_2$ to noise channel labeled $X$ as $V_X = G_X \delta X$ where $G_X$ is 
an operator on the Hilbert space spanned by $\gamma_1$ and $\gamma_2$. $|\delta X|^2$ is the ensemble average
value of the environmental noise sources as seen by the qubit.

The transition rate from an initial state $|\psi_i \rangle$ to a final state $|\psi_f \rangle$ can be calculated via Fermi's Golden Rule \cite{Groszkowski2018,Clerk2010,Fermi1932,Krantz2019}

\begin{equation}
\label{fermigolden}
  \Gamma_{i\rightarrow f}^{X,\pm} = \frac{1}{\hbar^2} \, |\langle \psi_f | G_X | \psi_i \rangle|^2 S_X ( |\omega_{fi}|).
\end{equation}

Here, the initial and final states are the ones established in Sec. II with energy difference $\hbar \omega_{fi} = E_f - E_i$. $S_X(\omega)$
is the noise spectral density of the noise source. The $\pm$ notation describes whether the rate is upwards ($\Gamma_{i \rightarrow f}^{X,+}$), where $E_f > E_i$, or downwards ($\Gamma_{i \rightarrow f}^{X,-}$), where $E_f < E_i$.
$G_X$ can also be considered as the qubit transverse susceptibility to
fluctuations $\delta X$ and is defined to be

\begin{equation}
  G_X = \frac{\partial \hat{H}}{\partial X}.
\end{equation}

This section is composed of subsections where each subsection derives the depolarization rate for different noise sources.

\subsection{\label{sub:ChargeDepolar}Charge Noise}

In order to calculate the charge noise effect on depolarization, first the charge noise
channel ($G_{n_{gi}}$) is defined in Eq. (\ref{Gngi}).

\begin{equation}
\label{Gngi}
  G_{n_{gi}}=\frac{\partial \hat{H}}{\partial n_{gi}} = 8   E_{C_i'}  (\hat n_i-n_{gi}) - 2E_{gi} n_{gi}
\end{equation}

The depolarization rate due to charge noise channel is calculated and demonstrated in
Fig.\ref{depolarization rates}a by assuming $A_{n_g}=10^{-4} 1/\sqrt{Hz}$ \cite{zorin1996background}.

As shown in Fig.\ref{depolarization rates}a, the depolarization rate due to
$n_{g1}$ (i.e. the biased side) contributes to the total rate more than $n_{g2}$ (i.e. the unbiased side)
since Degenerium's bias loop determines its characteristics. Moreover, similar to dephasing rate of the 
charge noise, depolarization rate due to charge noise is in a sweet spot which means Degenerium is relatively
resilient to charge noise fluctuations especially in the unbiased loop.
 
\subsection{Bias Flux Noise}
The depolarization rate due to bias flux noise is calculated similarly as in part A of Sec.\ref{sec:Depolarization} by assuming that the amplitude
of the 1/f flux noise  is $A_{\Phi}=10^{-6} \Phi_0/\sqrt{Hz}$ \cite{hutchings2017tunable}. Furthermore, the noise 
channel operator in Eq.(\ref{fermigolden}), assuming negligible junction energy difference, is calculated to be

\begin{equation}
\label{Gphibi}
  G_{\phi_{bi}}=\frac{\partial \hat{H}}{\partial \phi_{bi}} = 2  E_{Ji} \sin(\hat{\gamma_i} - \phi_{bi} - \phi_i/2) \cos(\phi_i/2).
\end{equation}

The bias flux noise, which is a crucial bottleneck of Degenerium, significantly determines the 
total depolarization rate of the qubit. However, the flux bias fluctuations do not worsen the 
depolarization rate unlike the pure dephasing one. The bias flux noise contributes 2 rad/s for each
loop to the total rate, which is lower than the corresponding pure dephasing rate.    
The results of simulation is shown in Fig.\ref{depolarization rates}b.

\subsection{Critical Current Noise}

To derive the depolarization rate caused by the critical current noise channel, first
we derive $G_{E_{Ji}}$ to be 

\begin{equation}
  G_{E_{Ji}}=\frac{\partial \hat{H}}{\partial E_{Ji}} = -2\cos(\hat{\gamma_i} - \phi_{bi} - \phi_i/2) \cos(\phi_i/2).
\end{equation}

Then we consider this channel to have a $1/f$ noise with the amplitude $A_{I_C}=10^{-7} I_C/\sqrt{Hz}$ \cite{Koch2007}. 
Thus, we employ Eq.(\ref{fermigolden}) and we calculate the depolarization rate for $E_{J1}$ and $E_{J2}$ which are depicted in 
Fig.\ref{depolarization rates}c.

As shown in the Fig.\ref{depolarization rates}c, the critical current noise has negligible effect on depolarization rate. The depolarization rate due to critical current noise is approximately $10^{-49}$ rad/s similar to its dephasing rate counterpart. Due to this feature, Degenerium is totally insensitive to critical current noise whether it is pure dephasing or depolarization rate.

\section{\label{sec:Conclusion}Conclusion}

In this work, a new qubit dubbed "Degenerium" is proposed which is inspired by QFP's topology and the 0-$\pi$ qubit concept.
Our Degenerium qubit features two ideally degenerate ground states which form the eigenbasis of the qubit. Unlike 0-$\pi$ qubit structure, Degenerium does not contain any inherently excessive harmonic modes, which has unwanted contributions to pure dephasing and depolarization rates.

Degenerium's potential can be engineered to host quadruple-well, double-well or even an
unbiased single-well potential, acting as a conventional flux qubit. Moreover, by setting
$E_{Ji}/E_{Li} \gg 1$, we can switch to a 0-$\pi$ qubit regime. All these choices make
Degenerium a versatile qubit that can adapt to designer's choice of operation and act as an 
application specific qubit.

The calculated pure dephasing and depolarization rates due to critical current noise are $3\times10^{-8}$ rad/s and  $1.4\times10^{-49}$ rad/s, respectively.

Additionally, The pure dephasing and depolarization rates due to charge noise are calculated to be $1.4\times10^3$ rad/s and 1 rad/s, respectively.

Furthermore, the calculated pure dephasing and depolarization rates due to flux noise are $7.1\times10^4$ rad/s and 4 rad/s, respectively.

The overall upper hand estimated $T_1$ (depolarization time) and $T_2$ (dephasing time) are obtained to be 1.25 s and 90 $\mu$s for the current choice of parameters of this work. Hence, the proposed Degenerium performs very well in terms of depolarization time ($T_1$), however, its decoherence time is currently limited to bias flux noise.
Further work can be performed for tuning Degenerium to enhance coherency.

\appendix

\section{\label{app:Hamiltonian}Derivation of Hamiltonian}
The Lagrangian described by Eq.(\ref{lagrangian}) consists of the kinetic term associated with the energy in the junction capacitances, the energy stored in the middle inductors, mutual inductive coupling and Josephson junctions.\\
The kinetic energy in the junction capacitances and the potential energy that stored in the inductors and Josephson junctions depend on the flux that passes through them.\\
The flux that passes through the left and right Josephson junction branch of Fig.\ref{fig:circuit} is calculated in Eq.(\ref{fluxleftandright}) with respect to the flux that passes through the middle inductors$(\Phi)$.The index $i\in \{1,2\}$ represents the left loop (for $i=1$) and right loop (for $i=2$) in $\Phi_{Li}$, $\Phi_{bi}$, $\Phi_{J_{i1}}$, $\Phi_{J_{i2}}$ and $\Phi_i$.
\begin{equation}
    \label{fluxleftandright}
    \begin{aligned}
      \Phi_{bi}&=-\Phi_{J_{i2}}+\Phi_{Li},
      &\Phi_{i}&=-\Phi_{J_{i1}}+\Phi_{J_{i2}} \\
    \end{aligned}
\end{equation}
The Lagrangian for left and right loop circuit is now described by Eq.(\ref{lagbefore}) which can be simplified according to Eq.(\ref{fluxleftandright}) to reach the final Eq.(\ref{lagrangian}).\\
\begin{equation}
    \label{lagbefore}
    \begin{aligned}
      \cal L &=\cal T-\cal U\\&=\frac{1}{2}C(\dot{\Phi}_{J_{i1}}^2+\dot{\Phi}_{J_{i2}}^2)\\&-\frac{1}{2L_i(1-K^2)}\Phi_{Li}^2\\
      &+(E_{Ji}-dE_{Ji})\cos(\frac{2\pi}{\Phi_0}(\Phi_{J_{i1}}))\\&+(E_{Ji}+dE_{Ji})\cos(\frac{2\pi}{\Phi_0}(\Phi_{J_{i2}}))\\&+\frac{M}{L_1L_2(1-K^2)}\Phi_{L_1}\Phi_{L_2}
    \end{aligned}
\end{equation}
where $K=M/\sqrt{L_1L_2}$. The conjugate momentum is defined as Eq.(\ref{conj}) considering the Lagrangian in Eq.(\ref{lagrangian}).
\begin{equation}
\label{conj}
  \begin{aligned}
    Q_i&=\frac{\partial \cal L}{\partial \dot{\Phi}_{Li}}=2C\dot{\Phi}_{Li}\\
    &
  \end{aligned}
\end{equation}
The classical Hamiltonian can be achieved as follows
\begin{equation}
\label{Ham}
  \begin{aligned}
    {\cal H}&=Q_i\dot{\Phi}_{Li}- {\cal L}=
    C\dot{\Phi}_{L_i}^2\\&+\frac{1}{2L_i(1-K^2)}\Phi_{Li}^2\\
      &-E_{Ji}\cos(\frac{2\pi}{\Phi_0}(\Phi_{Li}-\Phi_{bi}-\Phi_i/2))\cos(\frac{2\pi}{\Phi_0}(\Phi_i/2))
      \\&+dE_{Ji}\sin(\frac{2\pi}{\Phi_0}(\Phi_{Li}-\Phi_{bi}-\Phi_i/2))\sin(\frac{2\pi}{\Phi_0}(\Phi_i/2))\\&-\frac{M}{L_1L_2(1-K^2)}\Phi_{L_1}\Phi_{L_2}.
  \end{aligned}
\end{equation}
The obtained Hamiltonian in Eq.(\ref{Ham}) is simplified further by $\dot{\Phi}_{Li}=\frac{Q_i}{2C}$.\\
By $Q_i$ and $\Phi_{Li}$ promotion to operators and defining  number$(\hat{n}_i)$ and phase operator$(\hat{\gamma}_i)$ according to Eq.(\ref{nandphi}) with commutation relation described by Eq.(\ref{commu}), the final quantized Hamiltonian is achieved in Eq.(\ref{finalHam}).
\begin{equation}
    \label{nandphi}
    \hat{Q}_i=2e\hat{n}_i, 
    \hat{\Phi}_{Li}=\frac{\Phi_0}{2\pi}\hat{\gamma}_i
\end{equation}
\begin{equation}
    \label{commu}
    [\hat{\gamma}_i,\hat{n}_i]=i
\end{equation}
\begin{equation}
    \label{finalHam}
      \begin{aligned}
    \cal \hat{H}&= 4 \sum_{i=1}^{2} E_{C_i} \hat{n}_i^2 + \frac{1}{2} \sum_{i=1}^{2} E_{L_i} \hat{\gamma}_i^2 \\&- 2 \sum_{i=1}^{2} E_{J_i} \cos\left(\hat{\gamma}_i - \phi_{bi} - \frac{\phi_i}{2}\right) \cos\left(\frac{\phi_i}{2}\right)\\&+ 2 \sum_{i=1}^{2} dE_{J_i} \sin\left(\hat{\gamma}_i - \phi_{bi} - \frac{\phi_i}{2}\right) \sin\left(\frac{\phi_i}{2}\right)-g\hat{\gamma}_1\hat{\gamma}_2
  \end{aligned}
\end{equation}
where $E_{C_i}=\frac{e^2}{2C_{\Sigma_i}}$, $E_{Li}=\left(\frac{\Phi_0}{2\pi}\right)^2 \frac{1}{L_i (1 - K^2)}$, $E_{J_i} = \frac{\Phi_0}{2\pi} I_{c_i}$ and $g = \left(\frac{\Phi_0}{2\pi}\right)^2 \frac{M}{L_1 L_2 (1 - K^2)}$.
\section{\label{app:Dephasing}Derivation of Pure Dephasing Rate}
As mentioned in the main text, in presence of weak, uncorrelated, Gaussain noise channel which $\langle \delta X(t)\rangle=0$, the total Hamiltonian of the circuit can be written as:

\begin{equation}
H \approx H\left(X_{0}\right)+V_{X}(t)
\end{equation}

In the time dependent Schrodinger picture, the evolution of $|\psi(t)\rangle$ can be expressed as the basis of $H_0$ eigen basis ($|n\rangle$) or $|\psi(t)\rangle=c_n(t)|n\rangle$. Thus, by finding $c_n(t)$, the evolution of the total Hamiltonian can be solved.\\
The time evolution of $c_n(t)$ is described Schrodinger equation which is $i \frac{\mathrm{~d}}{\mathrm{~d} t} c_{n}(t)=\langle n|{V}_{X}(t)|{\psi}(t)\rangle=\sum_{n^{\prime}}\langle n| {V}_{X}(t)\left|n^{\prime}\right\rangle c_{n^{\prime}}(t)$. By assuming $V_X(t)$ only consists of diagonal terms that can be expanded as:

\begin{equation}
{V}_{X}(t)=\sum_{n} v_{n}(t)|n\rangle\langle n|
\end{equation}
Now the evolution of $|\psi(t)\rangle$ can be solved.

\begin{equation}
\left|{\psi}_{n}(t)\right\rangle=\exp (-\frac{i}{\hbar} \int_{0}^{t}  v_{n}(t^{\prime}) \mathrm{~d} t^{\prime} )|n\rangle 
\end{equation}
In order to find the estimated phase shift in $|\psi(t)\rangle$ due to the noise effect, we can find $v_n(t)$ by $H$ expand that we first introduced.

\begin{equation}
\begin{aligned}
    v_{n}(t)&=\langle n| \frac{\partial H}{\partial X}|_{X=X_0}|n\rangle \delta X(t)+\frac{1}{2}\langle n| \frac{\partial^2 H}{\partial X^2}|_{X=X_0}|n\rangle \delta X^{2}(t)\\&=\frac{\partial E_n}{\partial X}|_{X=X_0} \delta X(t)+\frac{\partial^2 E_n}{\partial X^2}|_{X=X_0} \delta X^{2}(t)
\end{aligned}
\end{equation}
Now, for computing pure dephasing rate which is defined the decay rate of diagonal density matrix element in Ramsey experiment, we can use below density matrix for initial condition of $|\psi(0)\rangle=1/\sqrt{2}(|0\rangle+|1\rangle)$.\\
\begin{equation}
    \rho(t)=\frac{1}{2}\left(\begin{array}{cc}
    1 & \rho_{01}(t)  \\
    \rho_{01}^{*}(t) & 1
    \end{array}\right) 
\end{equation}
where 
\begin{equation}
    \begin{aligned}
        \rho_{01}(t) &= \exp \bigg(-i \frac{\partial \omega_{01}}{\partial X}\bigg|_{X=X_0} \int_{0}^{t}  \delta X(t^{\prime}) \mathrm{d} t^{\prime} \\
        &\quad -i \frac{1}{2} \frac{\partial^2 \omega_{01}}{\partial X^2}\bigg|_{X=X_0} \int_{0}^{t}  \delta X^{2}(t^{\prime}) \mathrm{d} t^{\prime} \bigg)
    \end{aligned}
\end{equation}

For finding the decay rate, $\langle\rho_{01}(t)\rangle$ should be determined considering $\langle e^{iX}\rangle=e^{-\langle Y^2\rangle/2}$ for a Gaussian process. The final result in terms of noise power density($S_X(\omega)$) is shown in Eq.(\ref{rho01}).
\begin{equation}
    \label{rho01}
    \begin{aligned}
       & \langle\rho_{01}(t)\rangle = \exp \bigg[-\frac{1}{2} \left(\left.\frac{\partial\omega_{01}}{\partial X}\right|_{X_0}\right)^2 t^2 I_1  \\
        &-\frac{1}{4} \left(\left.\frac{\partial^2\omega_{01}}{\partial X^2}\right|_{X_0}\right)^2 \bigg(t^2 \sigma_X^4 + 2t^2 I_2 \bigg)\bigg]
    \end{aligned}
\end{equation}
where;

\begin{equation}
    \begin{aligned}
        I_1 &= \frac{1}{2\pi} \int  \operatorname{sinc}^2\left(\frac{\omega t}{2}\right) S_X(\omega) d\omega,
  \end{aligned}
\end{equation}
    
\begin{equation}
    \begin{aligned}
        I_2 &= \frac{1}{(2\pi)^2} \iint  \operatorname{sinc}^2\left(\frac{(\Omega+\omega)t}{2}\right) S_X(\omega)S_X(\Omega) d\Omega d\omega,       
\end{aligned}
\end{equation}
    
\begin{equation}
  \begin{aligned}
        \sigma_{X}^{2} &= \frac{1}{2 \pi} \int_{-\infty}^{\infty}  S_{X}(\omega) \mathrm{d} \omega
  \end{aligned}
\end{equation}

For $S_X(\omega)=2\pi A_X^2/|\omega|$ which is 1/f noise power density, decay rate or pure dephasing rate equivalently, will be

\begin{multline}
  \Gamma_X^\phi = 
  \Bigg[
  2 A_X^2 \left( \left. \frac{\partial \omega_{01}}{\partial X} \right|_{X=X_0} \right)^2 
  \left| \ln \left( \omega_{ir} \tau \right) \right| \\
  + 2 A_X^4 \left( \left. \frac{\partial^2 \omega_{01}}{\partial X^2} \right|_{X=X_0} \right)^2 
  \left( \ln^2 \left( \frac{\omega_c}{\omega_{ir}} \right) + 2 \ln^2 \left( \omega_{ir} \tau \right) \right)
  \Bigg]^{1/2}
\end{multline}.

\section{\label{app:Depolarization}Derivation of Charge noise Degenerium circuit Hamiltonian}
In case of charge noise which is modeled in Fig.\ref{fig:charge noise circuit}, The new Lagrangian is described by Eq.(\ref{chargenoiselag}).
\begin{equation}
    \label{chargenoiselag}
    {\cal L}_{n_{gi}}=\frac{1}{2}C_{gi}\dot{\Phi}_{Cgi}^2+\cal L
\end{equation}
where $\Phi_{Cgi}=\Phi_{Li}-\Phi_{gi}$ , $\Phi_{gi}$ and $\Phi_{Cgi}$ are the fluxes that passes through the virtual voltage source and coupling capacitance, respectively.\\
The conjugate momentum can be defined as,
\begin{equation}
     Q_{n_{gi}}=\frac{\partial {\cal L}_{n_{gi}}}{\partial \dot{\Phi}_{Li}}=(2C+C_{gi})\dot{\Phi}_{Li}-C_gV_g\\
\end{equation}
which will lead to Eq.(\ref{ngHam}) Hamiltonian.
\begin{equation}
    \label{ngHam}
    \begin{aligned}
         {\cal H}_{n_{gi}}&=Q_{n_{gi}}\Phi_{Li}-{\cal L}_{n_{gi}}=\frac{(Q_{n_{gi}}-C_{gi}V_{gi})^2}{2(2C+C_{gi})}\\&-\frac{(C_{gi}V_{gi})^2}{2C_{gi}}+\frac{1}{2L_i(1-K^2)}\Phi_{Li}^2\\
      &-E_{Ji}\cos(\frac{2\pi}{\Phi_0}(\Phi_{Li}-\Phi_{bi}-\Phi_i/2))\cos(\frac{2\pi}{\Phi_0}(\Phi_i/2))
      \\&+dE_{Ji}\sin(\frac{2\pi}{\Phi_0}(\Phi_{Li}-\Phi_{bi}-\Phi_i/2))\sin(\frac{2\pi}{\Phi_0}(\Phi_i/2))\\&-\frac{M}{L_1L_2(1-K^2)}\Phi_{L_1}\Phi_{L_2}
    \end{aligned}
\end{equation}
By following same approach as Appendix A, we arrive at quantized Hamiltonian that is described in Eq.(\ref{charge noise}).\\


\bibliography{references}

\end{document}